\documentclass[letter, twocolumn]{jpsj3}

\usepackage{color}
\usepackage{ulem}

\def\journal#1#2#3#4#5{#1: #2 {\bf #3} (#5) #4.}

\def\JPSJ{J.\ Phys.\ Soc.\ Jpn.}
\def\PR{Phys.\ Rev.}
\def\PRL{Phys.\ Rev.\ Lett.}
\def\PRB{Phys.\ Rev.\ B}

\def\RMP{Rev.\ Mod.\ Phys.}
\def\RPP{Rep.\ Prog.\ Phys.}

\def\JPC{J. Phys. C}
\def\EL{Europhys. Lett.}

\newcommand{\Ns}{\ensuremath{N_\mathrm{s}}}
\newcommand{\Ne}{\ensuremath{N_\mathrm{e}}}
\newcommand{\EF}{\ensuremath{E_\mathrm{F}}}



\def\runauthor{\textsc{H.~Shinaoka} \textit{et al.}}
\makeatletter
\def\@typeset{}
\makeatother

\begin{document}

\title{Electronic and Magnetic Properties of Metallic Phases under Coexisting Short-Range Interaction and Diagonal Disorder}
\author{Hiroshi {\sc SHINAOKA}$^{1,2}$\thanks{E-mail address: h.shinaoka@aist.go.jp} and Masatoshi {\sc IMADA}$^{2,3}$}
\inst{
$^{1}$ Nanosystem Research Institute, AIST, Tsukuba 305-8568\\
$^{2}$ CREST, JST, 7-3-1 Hongo, Bunkyo-ku, Tokyo 113-8656\\
$^{3}$ Department of Applied Physics, The University of Tokyo, Tokyo 113-8656} 
\recdate{\today}

\date{\today}
\abst{
We study a three-dimensional Anderson-Hubbard model under the coexistence of short-range interaction and diagonal disorder within the Hartree-Fock approximation.
We show that the density of states at the Fermi energy is suppressed in the metallic phases near the metal-insulator transition as a proximity effect of the soft Hubbard gap in the insulating phases. The transition to the insulator is characterized by a vanishing density of states (DOS) in contrast to the formation of a quasiparticle peak at the Fermi energy obtained using the dynamical mean field theory in pure systems. Furthermore, we show that there exist frozen spin moments in the paramagnetic metal.
}
\kword{soft Hubbard gap, electron correlation, disorder, Anderson-Hubbard model, single-particle density of states, pseudogap, Mott transition, Anderson localization, randomness, spin glass}

\maketitle

\section{Introduction}
Understanding the nature of metal-insulator transitions (MIT) has been a central issue in condensed matter physics for a long time~\cite{Imada98}. In particular, recently, the interplay of electron correlation and randomness has been attracting much attention experimentally and theoretically because of their inevitable coexistence in real materials.

The single-particle DOS is a typical physical quantity that characterizes not only the nature of the MITs but also the electronic and magnetic properties in their vicinity. When the electron correlation causes the MIT as Mott transition~\cite{Mott49}, it opens a gap in the single-particle DOS. On the other hand, the Anderson insulator, which is driven by randomness, exhibits no gap in the single-particle DOS~\cite{Anderson58}. Such contrasting behavior of the DOS for these two types of MITs raise naturally a simple question: How does the DOS behave near Mott-Anderson transitions with coexisting electron correlation and randomness? Despite extensive theoretical studies for several decades~\cite{criticality}, the nature of the Mott-Anderson transition has not yet been fully clarified.

The Anderson-Hubbard model with coexisting on-site repulsion and diagonal disorder is one of the simplest models suitable  for investigating the nature of the Mott-Anderson transition. Recently, we have determined the ground-state phase diagram of the three-dimensional Anderson-Hubbard model within the Hartree-Fock (HF) approximation. Furthermore, we have found an unconventional soft gap (\textit{soft Hubbard gap}) over the entire insulating phases~\cite{Shinaoka09a, Shinaoka09b}. Because only the short-range interaction is present in the Anderson-Hubbard model, the soft Hubbard gap cannot be explained by the conventional theory that attributes the formation of the soft gap to the long-range nature of the Coulomb interaction~\cite{Efros75}. Indeed, we have proposed a multivalley energy landscape, which may be characteristic of random systems, as the origin of the soft Hubbard gap. This observation of the soft gap is in clear contrast to the results of a numerical study within the dynamical mean-field theory (DMFT) ~\cite{Dobrosavljevic97} and some mean-field studies~\cite{Dobrosavljevic03, Byczuk05} that indicate the absence of the soft gaps. This contradiction may be due to the insufficient treatment of spatial correlation, which is essential in the formation of the soft gap, in those studies.

On the other hand, the behavior of the DOS in metals near MITs has been one of the central issues. In particular, \textit{pseudogap} phenomena observed in underdoped cuprate high-$T_\mathrm{c}$ superconductors have inspired fundamental discussions on the nature of the electronic states near the Mott insulator~\cite{Timusk99}. While the pseudogap and the soft gap both cause the reduction of the DOS at the Fermi level, the mechanism and origin of the pseudogap of the cuprates have not been established, and the role of randomness in stabilizing the superconductivity remains controversial. Recent numerical studies within the cellular DMFT indicates the stabilization of a pseudogap~\cite{Zhang07} or Fermi arc~\cite{Stanescu06a, Stanescu06b, Sakai09} near the MITs in pure systems, in contrast to the single-site DMFT results~\cite{Georges96}. Because the soft gap mechanism may deepen and constructively stabilize the pseudogap formation in real experimental circumstances with the inevitable coexistence of randomness and electron correlation, even on the HF level, further studies of the {\it metallic} phases near the Mott-Anderson transition will shed new light on the interplay of electron correlation and randomness and provide insight into the pseudogap phenomena.

Spin polarization (or the formation of frozen spin moments) is another essential element in determining magnetic properties in the vicinity of the MITs, such as uniform magnetic susceptibility. Although a previous HF study on the three-dimensional Anderson-Hubbard model claimed the formation of frozen spin moments even in the paramagnetic metal~\cite{Tusch93}, the analyses are limited to finite system sizes and the bulk limit was not analyzed after the extrapolation.

In this paper, we show further numerical analyses of single-particle excitations and spin polarization on the metallic side within the HF approximation.

\section{Model and Method}
The Anderson-Hubbard Hamiltonian is defined as
\begin{equation}
	\mathcal{H}=-t\sum_{\langle i,j \rangle,\sigma}c_{i\sigma}^{\dagger}c_{j\sigma} + U \sum_{i} n_{i \uparrow}n_{i \downarrow}+ \sum_{i, \sigma} (V_{i}-\mu) n_{i\sigma}
\end{equation}
on lattices with $\Ns$ sites and $\Ne$ electrons, where $t$ is the hopping integral, $U$ the on-site repulsion, $c_{i\sigma}^\dagger$ ($c_{i\sigma}$) the creation (annihilation) operator for an electron with spin $\sigma$ on site $i$, $n_{i\sigma}=c_{i\sigma}^\dagger c_{i\sigma}$, and $\mu$ is the chemical potential.
The random potential $V_i$ is spatially uncorrelated and assumed to follow the Gaussian type distribution $P_V(V_i)$ with the average $\langle V_i \rangle =0$: $P_V(V_i) =\frac{1}{\sqrt{2\pi} \sigma} \exp( -{V_i^2}/{2\sigma^2})$ ($\sigma^2 = W^2/12$). We focus on half filling ($\mu = U/2$) on the cubic lattice throughout this paper. We take the lattice spacing as the length unit.

We employ the Hartree-Fock (HF) approximation, where the wave function is approximated by a single Slater determinant consisting of a set of orthonormal single-particle orbitals $\{ \phi_n \}$ ($n$ is an orbital index). The HF equation reads
\begin{equation}
\{\mathcal{H}_0+U \sum_{i} ( \langle n_{i \downarrow} \rangle n_{i \uparrow}+ \langle n_{i \uparrow} \rangle n_{i \downarrow} ) \} \phi_n= \epsilon_n \phi_n,
\end{equation}
where $\mathcal{H}_0$ is the one-body part of the Hamiltonian and we neglect $\langle c_{i \uparrow}^\dagger c_{i \downarrow}\rangle$ and $\langle c_{i \downarrow}^\dagger c_{i \uparrow}\rangle$. To find a site-dependent mean-field solution $\langle n_{i \sigma} \rangle$ for the HF equations, we employ the iterative scheme. One typically needs from several to several tens of initial guesses to obtain convergent physical quantities such as antiferromagnetic (AF) order parameters and DOS. Here, we employ pseudo-one-dimensional unit cells of $L\times L\times M$, where $M \gg L$.

\section{Results: Electronic and Magnetic Properties of Paramagnetic Metal}
\subsection{Density of states}
\begin{figure}
 \centering
 \includegraphics[width=.4\textwidth,clip]{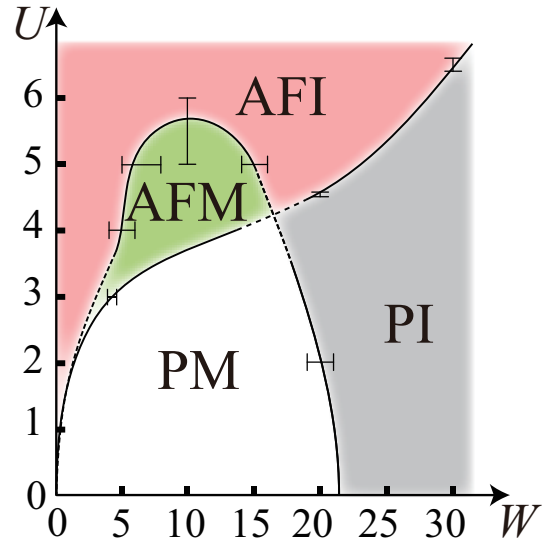}
 \caption{(Color online) Ground-state phase diagram of three-dimensional Anderson-Hubbard model at half filling for Gaussian distribution of $P_V$. AFI, AF insulator; AFM, AF metal; PI, paramagnetic insulator (Anderson insulator); PM, paramagnetic metal.}
 \label{fig:1}
\end{figure}

In Fig.~\ref{fig:1}, we present the calculated ground-state phase diagram within the HF approximation~\cite{Shinaoka09a, Shinaoka09b}. Hereafter, we take the hopping integral $t$ as the energy unit. At $U=0$, the Anderson-Hubbard model undergoes a metal-insulator transition (Anderson transition) from the paramagnetic metal (PM) to the paramagnetic insulator (PI) at a finite strength of disorder, $W_\mathrm{c}=21.29\pm0.02$~\cite{Slevin99}. On the other hand, at $W=0$, since the system is half-filled with perfect nesting, the ground state is the antiferromagnetic insulator (AFI) for any nonzero value of $U$. Here, we discuss the ground-state phase diagram for $U, W>0$. First, we focus on the spin degree of freedom. For $W>0$, the ground state is paramagnetic near $U=0$. With increasing interaction, the ground states undergo an antiferromagnetic transition at a critical point $U_\mathrm{c}~(>0)$. Within the resolution of our calculation, $U_\mathrm{c}$ monotonically increases as disorder strength $W$ increases. Next, we focus on the charge degrees of freedom. The ground state is insulating for $U, W\gg1$, which contains AFI as well as PI (PI is usually identified as an Anderson insulator). Metallic phases identified by the divergent localization length are restricted to a dome-like region ($U<6$ and $W<25$).

\begin{figure}
 \centering
 \includegraphics[width=.475\textwidth,clip]{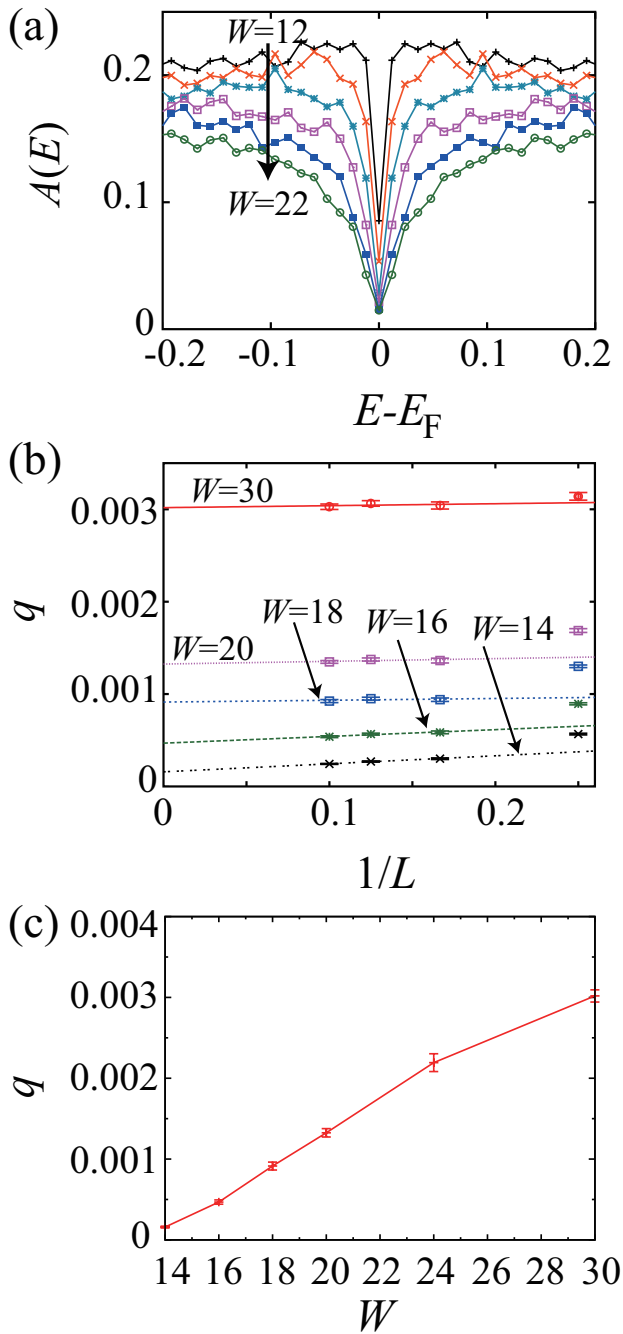}
 \caption{(Color online) (a) DOS with system size $8\times8\times250$ at $U=2$ ($W=22,~20,~18,~16,~14$, and $12$). We employ Lorentz broadening with a broadening factor of $5\times10^{-4}$. (b) The EA order parameter at $U=2$ as a function of inverse length $1/L$. (c) $W$ dependence of the EA order parameter extrapolated to the bulk limit at $U=2$.}
 \label{fig:2}
\end{figure}

Now, we discuss the DOS in the paramagnetic metal. Figure~\ref{fig:2} (a) shows the DOS at $U=2$  ($12 \le W \le 22$) across the MIT point. At $U=0$, the DOS shows no anomaly in the paramagnetic metal even near the MIT, because the DOS is not an order parameter of the Anderson transition and $A(E_\mathrm{F})$ is nonzero in the Anderson insulator. In contrast, for $U>0$, the DOS at the Fermi energy should be zero in the paramagnetic insulator, because of the formation of the soft Hubbard gap~\cite{Shinaoka09a, Shinaoka09b}. Indeed, the DOS shows a dip even in the metallic phase for $12 \le W \le 20$ as a result of the proximity effect of the soft Hubbard gap. The dip gradually becomes deeper with increasing $W$, and finally turns into the soft gap at the MIT. We have confirmed that the DOS for $L=8$ and $L=10$ are not different within the error bars at $W=14$, indicating that the reduction of the DOS remains in the bulk limit. This critical behavior of the DOS is opposite to the results of a previous numerical study within the framework of the dynamical mean-field theory (DMFT) in infinite spatial dimensions, which support the divergence of $A(\EF)$ toward the MIT~\cite{Dobrosavljevic97}. This suppression of the DOS in our HF study indicates that the physical properties of the paramagnetic metal for $U>0$ is different from those of the non-interacting paramagnetic metal. For example, the Fermi surface in the paramagnetic metal vanishes as the dip becomes deeper toward the MIT. Because the momentum-resolved DOS can be observed directly by angle-resolved photoemission spectroscopy, how the Fermi surface vanishes in the momentum space is an interesting problem left for future studies. Furthermore, because the magnetic susceptibility is proportional to the DOS at the Fermi energy, the Pauli susceptibility should be strongly reduced toward the MIT. However, it should be noted that the clarification of contributions from the incoherent part is needed to reach the complete understanding of the magnetic properties.

\subsection{Spin polarization/formation of frozen spin moments}
Next, we discuss spin degrees of freedom contained also in the incoherent part. The Edwards-Anderson (EA) order parameter for the spin glass is given by $q \equiv 4\Ns^{-1}\sum_{i} {\langle S_i \rangle}^2$~\cite{Edwards75}. Figure~\ref{fig:2}(b) shows an extrapolation of the EA order parameter to the bulk limit.
In fact, we adopt $M=1000, 250, 250$, and $100$ for $L=4, 6, 8$, and $10$, respectively. We find that the EA order parameter is extrapolated to a nonzero value even in the thermodynamic limit of the paramagnetic metal region, indicating the existence of frozen moments arising from the coexistence of the electron correlation and randomness. As shown in Fig.~\ref{fig:2}(c), the EA order parameter extrapolated to the bulk limit exhibits no singularity at the MIT. This formation of frozen moments in the paramagnetic metal is consistent with the result of a previous study using an effective-field theory~\cite{Milovanovic89}. 
However, the present HF calculations break the SU(2) symmetry and ignore quantum fluctuations.
We need further studies beyond the mean-field level toward a complete understanding of the magnetic properties in the paramagnetic metal.

\section{Summary and Future Perspective}
We analyzed the three-dimensional Anderson-Hubbard model within the HF approximation. We found a dip in the DOS centered at the Fermi energy in metallic phases near the MIT for $U>0$. The transition from a metal to an Anderson-Mott insulator is characterized by a continuously vanishing DOS in metals at the Fermi level, in sharp contrast to the picture of the dynamical mean field theory. Furthermore, we observed the formation of frozen moments in the paramagnetic metal with coexisting electron correlation and randomness, which may be essential in determining the magnetic properties.

In the previous papers~\cite{Shinaoka09a, Shinaoka09b}, as the origin of the soft Hubbard gap, we proposed the existence of low-energy multiply-excited states with electronic structures that are globally relaxed from those of the ground state. There exist many excited states nearly degenerated with the ground state generically in a spin-glass phase. Therefore it will be interesting to clarify the connection between the spin-glass freezing and the formation of the soft Hubbard gap in the Anderson-Hubbard model.

Before closing this paper, we make a brief remark concerning future topics. The present HF results support the stabilization of a pseudogap in a metal by the interplay of electron correlation and randomness, which may play a certain role in the formation of the pseudogap in various materials such as cuprates~\cite{Norman97} and Ca$_{1-x}$Sr$_x$VO$_3$~\cite{Eguchi06}. In the context of the Fermi arc observed in photoemission for the cuprates~\cite{Norman97}, it will be of great interest to investigate momentum-resolved DOS in a metal using a two-dimensional Anderson-Hubbard model.

\begin{acknowledgments}
	Numerical calculation was partly carried out at the Supercomputer Center, Institute for Solid State Physics, Univ. of Tokyo. This work is financially supported by MEXT under grant number 17071003. 
\end{acknowledgments}

\end{document}